\documentclass[twocolumn,showkeys,aps,prb,showpacs]{revtex4-1}
\usepackage{graphicx}
\usepackage{amsmath}
\usepackage[CJKbookmarks,dvipdfm,colorlinks,linkcolor=blue,citecolor=blue]{hyperref}

\begin{document}

\title{Phonon transport of Janus monolayer MoSSe: a first-principles study}
\author{San-Dong Guo}
\affiliation{School of Physics, China University of Mining and
Technology, Xuzhou 221116, Jiangsu, China}

\begin{abstract}
Transition Metal Dichalcogenide (TMD) monolayers  have most widely studied  due to
their unique physical properties. Recently, Janus TMD Monolayer MoSSe with sandwiched S-Mo-Se structure has been synthesized by replacing the top S atomic layer in  $\mathrm{MoS_2}$ with Se atoms. In this work, we systematically investigate the
phonon transport and lattice  thermal conductivity ($\kappa_L$) of  MoSSe monolayer  by  first-principles calculations and  linearized phonon Boltzmann equation within the single-mode relaxation time approximation (RTA). 
Calculated results show that the $\kappa_L$ of MoSSe monolayer is very lower than that of $\mathrm{MoS_2}$ monolayer, and higher than that of  $\mathrm{MoSe_2}$ monolayer. The corresponding sheet thermal conductance of MoSSe monolayer is  342.50  $\mathrm{W K^{-1}}$  at room temperature.
These can be understood by phonon group velocities and  lifetimes. Compared with $\mathrm{MoS_2}$ monolayer,  the smaller group velocities  and shorter phonon lifetimes of  MoSSe monolayer give rise to lower $\kappa_L$.  The larger group velocities for MoSSe than $\mathrm{MoSe_2}$ monolayer is main reason of higher $\kappa_L$.  The elastic properties of $\mathrm{MoS_2}$, MoSSe and $\mathrm{MoSe_2}$ monolayers are also calculated, and the order of Young's modulus  is identical with that of $\kappa_L$. Calculated results show that isotope scattering  leads to 5.8\% reduce of $\kappa_L$.  The  size effects on the $\kappa_L$ are also considered, which is usually used  in the device implementation.  When the characteristic length of MoSSe monolayer  is about 110 nm,  the $\kappa_L$  reduces  to half. These results may offer perspectives on thermal management of MoSSe monolayer  for applications of thermoelectrics, thermal
circuits and nanoelectronics, and motivate further theoretical or experimental efforts to investigate thermal transports of  Janus TMD monolayers.
\end{abstract}

\pacs{72.15.Jf, 71.20.-b, 71.70.Ej, 79.10.-n ~~~~~~~~~~~~~~~~~~~~~~~~~~~~~~~~~~~Email:sandongyuwang@163.com}

\maketitle

\section{Introduction}
Due to many novel properties, two-dimensional (2D) materials have been
attracting increasing attention since the discovery of graphene\cite{q1}.
The TMD\cite{q7}, group-VA\cite{q9,q10}, group IV-VI\cite{q8} and group-IV\cite{q11}  monolayers have been predicted theoretically or synthesized experimentally, which have  potential applications in  electronic,  thermoelectric, quantum and optoelectronic devices.
Recently,  Janus monolayer MoSSe has been synthesized, based on $\mathrm{MoS_2}$ monolayer by  breaking the out-of-plane structural
symmetry\cite{p1}. The existence of vertical
dipoles  has been proved  by second harmonic generation and piezoresponse
force microscopy measurements\cite{p1}.  The  strong piezoelectric effects have been predicted in
monolayer and multilayer Janus TMD  MXY (M = Mo and  W; X/Y = S, Se and Te) by first-principles calculations\cite{p2}, which has potential applications in energy harvesting and  sensors. Electronic and optical properties have been studied  in pristine Janus MoSSe and WSSe monolayers and their  vertical and lateral heterostructures\cite{p3}. The  ZrSSe monolayer has also been predicted with  the 1T phase\cite{p4},  which is different from  MoSSe monolayer with 2H phase. It is proved that ZrSSe monolayer has better n-type thermoelectric properties than monolayer  $\mathrm{ZrS_2}$.

\begin{figure}
  \includegraphics[width=5.5cm]{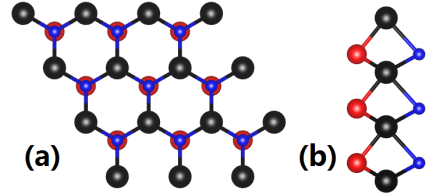}
  \caption{(Color online) Top (Right) and side (Left) views of  crystal structure
of MoSSe monolayer.  The large black balls represent Mo atoms, and the middle  red balls for Se atoms,  and the small  blue balls for  S atoms.}\label{st}
\end{figure}
\begin{figure*}
  \includegraphics[width=15cm]{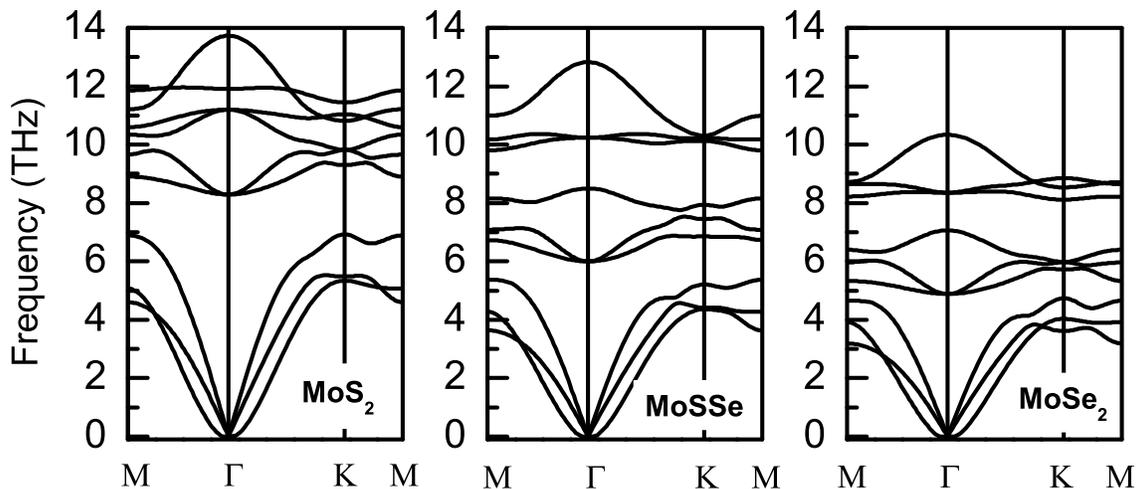}
  \caption{Phonon dispersion curves of $\mathrm{MoS_2}$, MoSSe and $\mathrm{MoSe_2}$ monolayers.}\label{ph}
\end{figure*}

 The thermal property of 2D materials  is quite worth studying due to its importance on the performance
and reliability on the nano-devices. As is well known,  a high thermal conductivity can effectively remove the accumulated heat, while  a low thermal conductivity is beneficial to thermoelectric applications.
In theory, thermal transports of many 2D materials  have been widely studied\cite{p5,q21,q22,l1,l2,l4,l4-1,l4-2,l4-3,l7,l71,l8,l9,l10,l102,l100,l101}, such as TMD, group-VA, ATeI (A=Sb or Bi), group IV-VI and group-IV monolayers. It is found that strain can effectively tune $\kappa_L$ for various kinds of 2D materials, such as group-IV monolayers\cite{l9,l101,l100},  $\mathrm{MoTe_2}$\cite{l4-3}, antimonene\cite{l102} and  Penta-Structures monolayers\cite{l10}.
With  strain increasing, the $\kappa_L$ shows monotonous increase/decrease  and  up-and-down behavior, and  tensile strain can induce strong size effects on $\kappa_L$.   The phonon transports of  TMD $\mathrm{MX_2}$ (M=Mo, W, Zr and Hf; X=S and Se) monolayers have been systematically studied by  phonon Boltzmann
transport equation approach\cite{p5}. The $\kappa_L$
of 2H-type TMD monolayers  are generally higher than those of  1T-type
ones, which can be attributed to the large acoustic-optical frequency gap\cite{p5}.
In this work, the  phonon transport of Janus TMD MoSSe monolayer is performed  from a combination of  first-principles calculations and  linearized phonon Boltzmann equation.  It is found that  the $\kappa_L$ of MoSSe monolayer  is very  lower than that  of $\mathrm{MoS_2}$ monolayer, but higher than one of $\mathrm{MoSe_2}$ monolayer.
 The order of their $\kappa_L$ is explored by  phonon group velocities and lifetimes.  It is found that  the order of Young's modulus ($\mathrm{MoS_2}$ $>$ MoSSe $>$  $\mathrm{MoSe_2}$) is identical with that of $\kappa_L$, which accords with the relation: $\kappa_L\sim \sqrt{E}$\cite{q16}.
  The isotope and size effects on  $\kappa_L$  are also studied, which can  provide  valuable information  for designing MoSSe-based nano-electronics devices.

The rest of the paper is organized as follows. In the next
section, the computational details about   phonon transport calculations are given. In the third section, the  phonon transport and elastic properties of MoSSe monolayer, together with ones of $\mathrm{MoS_2}$ and $\mathrm{MoSe_2}$ monolayers for a comparison, are shown. Finally, we shall give some discussions and conclusions in the fourth section.
\begin{table}
\centering \caption{Lattice constants $a$  and  related bond lengths of $\mathrm{MoS_2}$, MoSSe and $\mathrm{MoSe_2}$ monolayers in $\mathrm{{\AA}}$. }\label{tab}
  \begin{tabular*}{0.48\textwidth}{@{\extracolsep{\fill}}cccccc}
  \hline\hline
Name& $a$ & $d_{Mo-S}$ &   $d_{Mo-Se}$&  $d_{S/Se-S/Se}$       \\\hline\hline
$\mathrm{MoS_2}$&   3.18 (3.17\cite{p1})   &   2.41  &  -&3.12 \\\hline
MoSSe&   3.25  (3.23\cite{p1}) &    2.42     &   2.54 &3.23    \\\hline
$\mathrm{MoSe_2}$&  3.30 (3.30\cite{p1})      &   -  &  2.54 &3.35      \\\hline\hline
\end{tabular*}
\end{table}

\begin{table*}
\centering \caption{The calculated optical phonon frequencies (THz) of $\mathrm{MoS_2}$, MoSSe and $\mathrm{MoSe_2}$ monolayers at
the $\Gamma$ point with  experimental results given in parentheses.}\label{tab1}
  \begin{tabular*}{0.96\textwidth}{@{\extracolsep{\fill}}cccccc}
  \hline\hline
Name&  $\mathrm{E}^{"}$& $\mathrm{E}^{'}$ &  $\mathrm{A}^{'}_1$&  $\mathrm{A}^{"}_2$       \\\hline\hline
$\mathrm{MoS_2}$&   8.29 (8.49\cite{p6})  &   11.20 (11.55\cite{p6}) &  11.91 (12.12\cite{p6}) &13.73 (14.10\cite{p6}) \\\hline
MoSSe&   6.00   &  10.24 (10.65\cite{p1}) &  8.50 (8.64\cite{p1})    &   12.83    \\\hline
$\mathrm{MoSe_2}$&  4.88  (5.01\cite{p7})    & 8.35(8.46\cite{p7}) &  7.06 (7.20\cite{p7}) &  10.34 (10.53\cite{p7})    \\\hline\hline
\end{tabular*}
\end{table*}

\section{Computational detail}
 Within projector augmented-wave method, we perform the first-principles calculations using the VASP code\cite{pv1,pv2,pv3,pbe} by adopting
 generalized gradient approximation   of Perdew-Burke-Ernzerhof (PBE-GGA)  as exchange-correlation functional\cite{pbe}.
During structural relaxation,  a 20 $\times$ 20 $\times$ 1 k-mesh is used  with  a Hellman-Feynman force convergence threshold of $10^{-4}$ eV/ $\mathrm{{\AA}}$. A plane-wave basis set is employed with
kinetic energy cutoff of 450 eV, and  the electronic stopping criterion is $10^{-8}$ eV. 
The 5s and 4d electrons of Mo,  and  3/4s and 3/4p electrons of S/Se  are treated as valance ones.
The  lattice thermal conductivity is performed
by using Phono3py+VASP codes\cite{pv1,pv2,pv3,pv4}.
By solving linearized phonon Boltzmann equation, the $\kappa_L$ is calculated with single-mode RTA,  as implemented in the Phono3py code\cite{pv4}. The $\kappa_L$ can be expressed as:
\begin{equation}\label{eq0}
    \kappa=\frac{1}{NV_0}\sum_\lambda \kappa_\lambda=\frac{1}{NV_0}\sum_\lambda C_\lambda \nu_\lambda \otimes \nu_\lambda \tau_\lambda
\end{equation}
in which  $\lambda$, $N$ and  $V_0$ are  phonon mode, the total number of q points sampling Brillouin zone (BZ) and  the volume of a unit cell, and  $C_\lambda$,  $ \nu_\lambda$, $\tau_\lambda$   is the specific heat,  phonon velocity,  phonon lifetime.
The phonon lifetime $\tau_\lambda$ can be attained  by  phonon linewidth $2\Gamma_\lambda(\omega_\lambda)$ of the phonon mode
$\lambda$:
\begin{equation}\label{eq0}
    \tau_\lambda=\frac{1}{2\Gamma_\lambda(\omega_\lambda)}
\end{equation}
The $\Gamma_\lambda(\omega)$  takes the form analogous to the Fermi golden rule:
\begin{equation}
\begin{split}
   \Gamma_\lambda(\omega)=\frac{18\pi}{\hbar^2}\sum_{\lambda^{'}\lambda^{''}}|\Phi_{-\lambda\lambda^{'}\lambda^{''}}|^2
   [(f_\lambda^{'}+f_\lambda^{''}+1)\delta(\omega
    -\omega_\lambda^{'}-\\\omega_\lambda^{''})
   +(f_\lambda^{'}-f_\lambda^{''})[\delta(\omega
    +\omega_\lambda^{'}-\omega_\lambda^{''})-\delta(\omega
    -\omega_\lambda^{'}+\omega_\lambda^{''})]]
\end{split}
\end{equation}
in which $f_\lambda$  and $\Phi_{-\lambda\lambda^{'}\lambda^{''}}$ are the phonon equilibrium occupancy and the strength of interaction among the three phonons $\lambda$, $\lambda^{'}$, and $\lambda^{''}$ involved in the scattering.
Based on the supercell approach  with finite atomic displacement
of 0.03 $\mathrm{{\AA}}$,  the second-order interatomic force constants (IFCs) can be attained
by using   a 5 $\times$ 5 $\times$ 1   supercell   with k-point meshes of 2 $\times$ 2 $\times$ 1.
According to second-order  harmonic IFCs, phonon dispersions can be calculated by  Phonopy package\cite{pv5}.  The third-order  IFCs can be attained
by using   a 3 $\times$ 3 $\times$ 1  supercell   with k-point meshes of 3 $\times$ 3 $\times$ 1.
To compute accurately lattice thermal conductivity,  the reciprocal spaces of the primitive cells  are sampled by 100 $\times$ 100 $\times$ 1 meshes .

For 2D material, the calculated  lattice thermal conductivity  depends on the length of unit cell along z direction\cite{2dl}.  They  should be normalized by multiplying $Lz/d$, where  $Lz$ and $d$ are the length of unit cell along z direction and  the thickness of 2D material.  However, the $d$  is not well defined,  for example graphene.  In this work, the $Lz$=24.64 $\mathrm{{\AA}}$  is used as $d$. By $\kappa$ $\times$ $d$,  the thermal sheet conductance can be attained.
\begin{figure}
  \includegraphics[width=8cm]{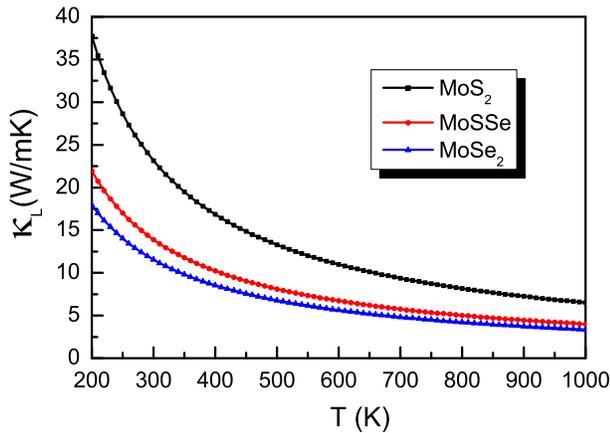}
  \caption{(Color online)The lattice thermal conductivities of $\mathrm{MoS_2}$, MoSSe and $\mathrm{MoSe_2}$ monolayers as a function of temperature.}\label{kl}
\end{figure}

\section{MAIN CALCULATED RESULTS AND ANALYSIS}
The structure of Janus MoSSe monolayer  is similar to  $\mathrm{MoS_2}$/$\mathrm{MoSe_2}$  monolayer with the 2H phase, containing  three atomic sublayers with Mo layer sandwiched between S and Se layers.
The  Janus monolayer MoSSe can be constructed  by  replacing one of two  S (Se)  layers with Se (S)  atoms in  $\mathrm{MoS_2}$ ($\mathrm{MoSe_2}$) monolayer.
The schematic crystal structure of MoSSe monolayer is plotted in \autoref{st}. It is clearly seen that the Janus MoSSe
monolayer loses the reflection symmetry with respect to the central metal Mo atoms compared with $\mathrm{MoS_2}$/$\mathrm{MoSe_2}$ monolayer. Therefore, the MoSSe monolayer (No.156) has lower symmetry compared with $\mathrm{MoS_2}$/$\mathrm{MoSe_2}$ monolayer (No.187).
To avoid spurious interaction between neighboring layers, the unit cell  of  Janus MoSSe monolayer,  containing  one Mo, one S and one Se atoms, is constructed with the vacuum region of more than 18 $\mathrm{{\AA}}$. The optimized lattice constants (other theoretical values\cite{p1}) and bond lengths of $\mathrm{MoS_2}$, MoSSe and $\mathrm{MoSe_2}$ monolayers are listed in \autoref{tab}.
It is expected that $a$ of MoSSe monolayer is between ones of   $\mathrm{MoS_2}$ and   $\mathrm{MoSe_2}$ monolayers,
which is about 2.2\%  higher than that of  $\mathrm{MoS_2}$ monolayer, and 1.5\% lower than that of  $\mathrm{MoSe_2}$ monolayer.
It is noted  that the bond length of Mo-S/Se between MoSSe and $\mathrm{MoS_2/Se_2}$ monolayers is almost the same. The bond length of S-Se of MoSSe monolayer is between ones of S-S ($\mathrm{MoS_2}$) and Se-Se ($\mathrm{MoSe_2}$).

\begin{figure}
  \includegraphics[width=8cm]{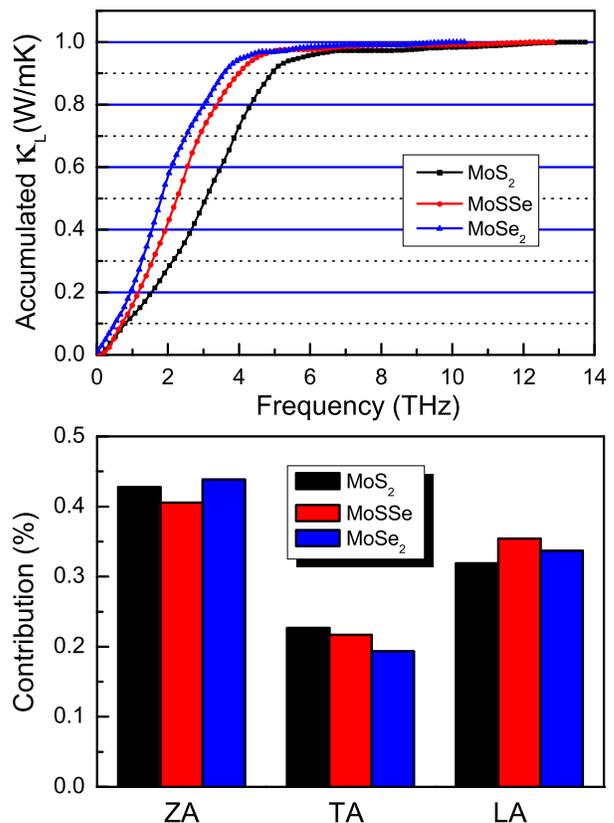}
  \caption{(Color online)For $\mathrm{MoS_2}$, MoSSe and $\mathrm{MoSe_2}$ monolayers: Top: the ratio between  accumulated and  total lattice thermal conductivity with respect to frequency. Bottom: the phonon modes contributions toward total lattice thermal conductivity from ZA, TA and LA acoustic phonon branches.}\label{mode}
\end{figure}

\autoref{ph} shows the phonon dispersions of $\mathrm{MoS_2}$, MoSSe and $\mathrm{MoSe_2}$ monolayers  along  high symmetry path, which agree well with previous results\cite{p1,p5,l4}.  The 3 acoustic
and 6 optical phonon branches are observed due to three atoms in the unit cell.
The longitudinal acoustic (LA) and transversal acoustic (TA) branches are linear near the $\Gamma$ point, while out-of-plane acoustic (ZA) branch  deviates from linearity.  Similar behavior can be found in many 2D materials\cite{q21,q22,l1,l2,l4,l7,l71,l8,l9,l10,l102,l100,l101}.
Due to  $D_{3h}$ symmetry for $\mathrm{MoS_2}$, MoSSe and $\mathrm{MoSe_2}$ monolayers,  the optical
lattice-vibration modes at $\Gamma$ point can be  defined as:
\begin{equation}\label{e1}
\Gamma_{optical}\equiv A^{"}_2(IR)+A^{'}_1(R)+E^{'}(IR+R)+E^{"}(R)
\end{equation}
in which  IR and R mean  infrared- and Raman-active mode, respectively.
The optical phonon frequencies of $\mathrm{MoS_2}$, MoSSe and $\mathrm{MoSe_2}$ monolayers at the
$\Gamma$ point along with available experimental values  are  listed in \autoref{tab1}. The calculated phonon frequencies of $\mathrm{MoS_2}$ and $\mathrm{MoSe_2}$ monolayers are in agreement with the experimental results\cite{p1,p6,p7}.
From  $\mathrm{MoS_2}$ to  MoSSe to $\mathrm{MoSe_2}$ monolayer,  acoustic modes  become softened,
and the optical branches overall move toward lower energy, which mean reduced group velocities. A frequency gap
between the  acoustic and optical phonon branches  can be observed, which is due to mass differences between the constituent atoms\cite{m1-1,m3-1}.
The frequency gap  is 1.36 THz for $\mathrm{MoS_2}$,   0.63 THz for MoSSe and   0.15 THz for $\mathrm{MoSe_2}$.  The frequency gap along with the width of acoustic branches are listed in \autoref{tab2}, which agree well with available theoretical results\cite{p5,l4}.
It is noted that the frequency gap can produce important influence on acoustic+acoustic$\rightarrow$optical (aao) scattering\cite{p5}.  The large gap  induces ineffective aao scattering due to the requirement on energy conservation, while small gap results in much more frequent aao
scattering. These have important effects on phonon transports of both bulk and 2D materials\cite{h5,h6,p5}.

\begin{figure}
  \includegraphics[width=8cm]{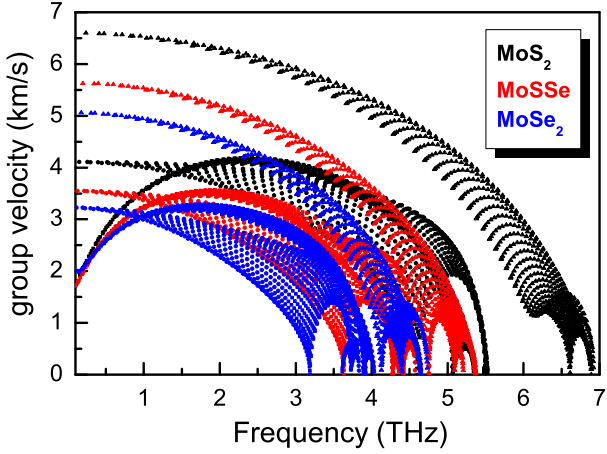}
  \caption{(Color online) The phonon mode group velocities of  $\mathrm{MoS_2}$ (Black), MoSSe (Red) and $\mathrm{MoSe_2}$ (Blue) monolayers in the first BZ  for ZA (square symbol), TA (circle symbol) and LA (UpTriangle symbol) acoustic branches. }\label{v}
\end{figure}
\begin{figure}
  \includegraphics[width=8cm]{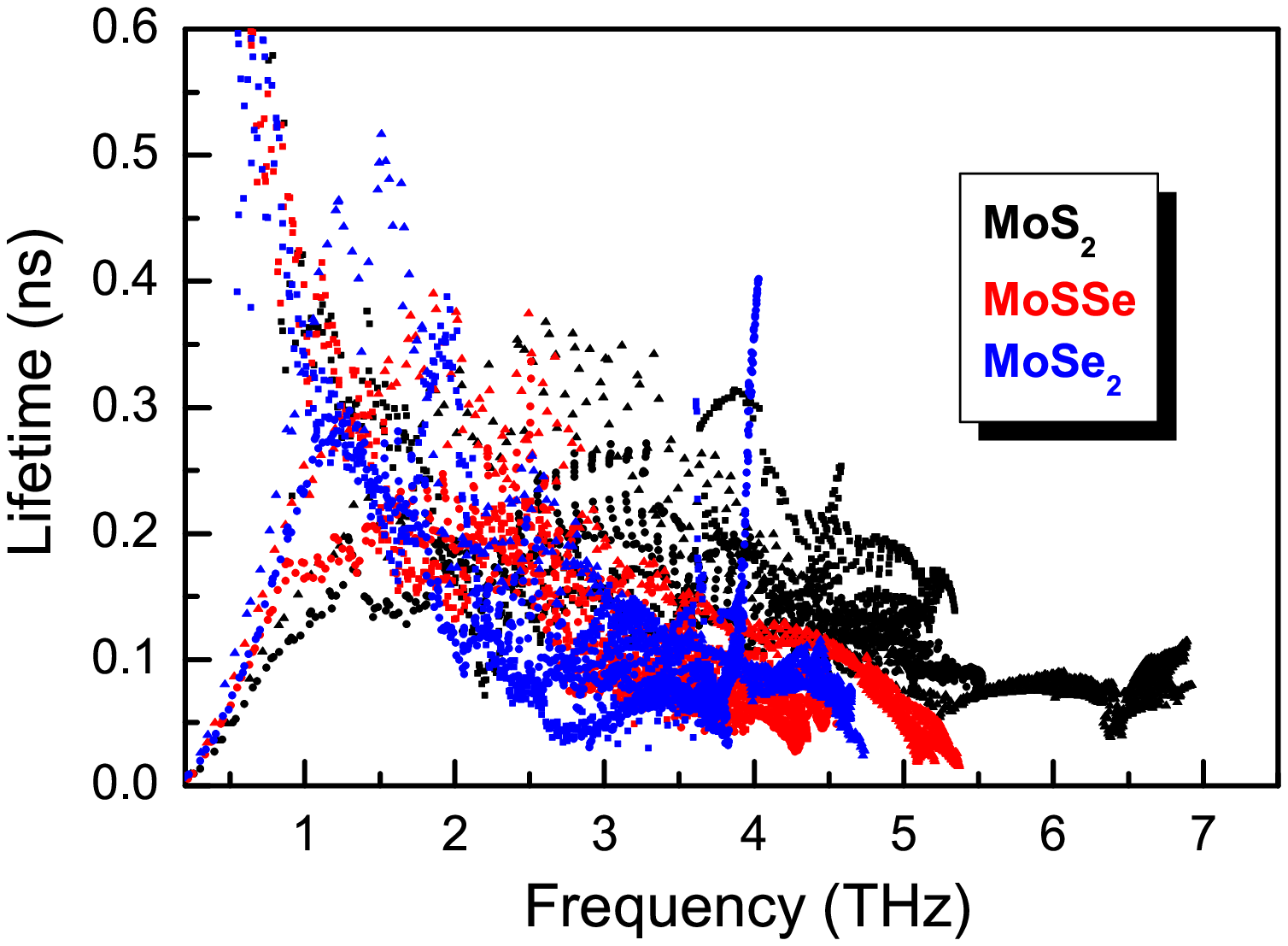}
  \caption{(Color online) The phonon mode lifetimes  of  $\mathrm{MoS_2}$ (Black), MoSSe (Red) and $\mathrm{MoSe_2}$ (Blue) monolayers in the first BZ for ZA (square symbol), TA (circle symbol) and LA (UpTriangle symbol) acoustic branches.}\label{t}
\end{figure}

\begin{table}
\centering \caption{The frequency gap between the  acoustic and optical phonon branches $G_{ao}$ (THz); the width of acoustic branches $W_a$ (THz);  thermal sheet conductance $\kappa_L$ ($\mathrm{W K^{-1}}$). }\label{tab2}
  \begin{tabular*}{0.48\textwidth}{@{\extracolsep{\fill}}ccccc}
  \hline\hline
Name& $G_{ao}$ & $W_a$ &  $\kappa_L$\\\hline\hline
$\mathrm{MoS_2}$&1.36 (1.35\cite{p5})&6.93 (6.90\cite{p5},7.01\cite{l4})   &570.42 (509.84\cite{p5})   \\\hline
MoSSe&0.63 &5.37   & 342.50\\\hline
$\mathrm{MoSe_2}$& 0.15 &4.74 (4.73\cite{l4})  &284.35 (307.33\cite{p5})\\\hline\hline
\end{tabular*}
\end{table}

Within RTA method,  \autoref{kl} shows the  intrinsic lattice  thermal conductivities of $\mathrm{MoS_2}$, MoSSe and $\mathrm{MoSe_2}$ monolayers from harmonic and anharmonic IFCs.  With the same thickness $d$ (24.64 $\mathrm{{\AA}}$), the room-temperature lattice  thermal conductivity  is 23.15 $\mathrm{W m^{-1} K^{-1}}$, 13.90 $\mathrm{W m^{-1} K^{-1}}$  and 11.54 $\mathrm{W m^{-1} K^{-1}}$, respectively. Their  thermal sheet conductance\cite{2dl} is  570.42 $\mathrm{W K^{-1}}$,  342.50  $\mathrm{W K^{-1}}$  and  284.35 $\mathrm{W K^{-1}}$, respectively. The thermal sheet conductances  of $\mathrm{MoS_2}$, MoSSe and $\mathrm{MoSe_2}$ monolayers are listed in \autoref{tab2}, together with reported theoretical values\cite{p5} using similar RTA method, which have been converted into thermal sheet conductances.  Our calculated values of  $\mathrm{MoS_2}$ and $\mathrm{MoSe_2}$ monolayers  are  very close to previous ones\cite{p5}.
It is expected that the  lattice  thermal conductivity of MoSSe monolayer is  between ones of $\mathrm{MoS_2}$ and $\mathrm{MoSe_2}$ monolayers.
In the considered  temperature range, the $\kappa_L$ of MoSSe monolayer is about 60\% of one of  $\mathrm{MoS_2}$ monolayer, and around  121\% of $\kappa_l$ of $\mathrm{MoSe_2}$. For $\mathrm{MoS_2}$, MoSSe and $\mathrm{MoSe_2}$ monolayers, the ratio between  accumulated and  total lattice thermal conductivity with respect to frequency  are  plotted in \autoref{mode} at room temperature. It is clearly seen that acoustic branches of  $\mathrm{MoS_2}$, MoSSe and $\mathrm{MoSe_2}$ monolayers dominate lattice  thermal conductivity,  providing  a contribution of 97.3\%, 97.6\% and 96.9\%, respectively.
 The relative contribution of every phonon  mode of acoustic branches  to the total lattice thermal conductivity (300 K) also are shown in \autoref{mode}.
It is found that the order of  contribution is ZA $>$ LA $>$ TA for all three monolayers, and about 42\% for ZA mode, 33\% for TA mode and 21\% for LA mode.
\begin{figure}
  \includegraphics[width=8.0cm]{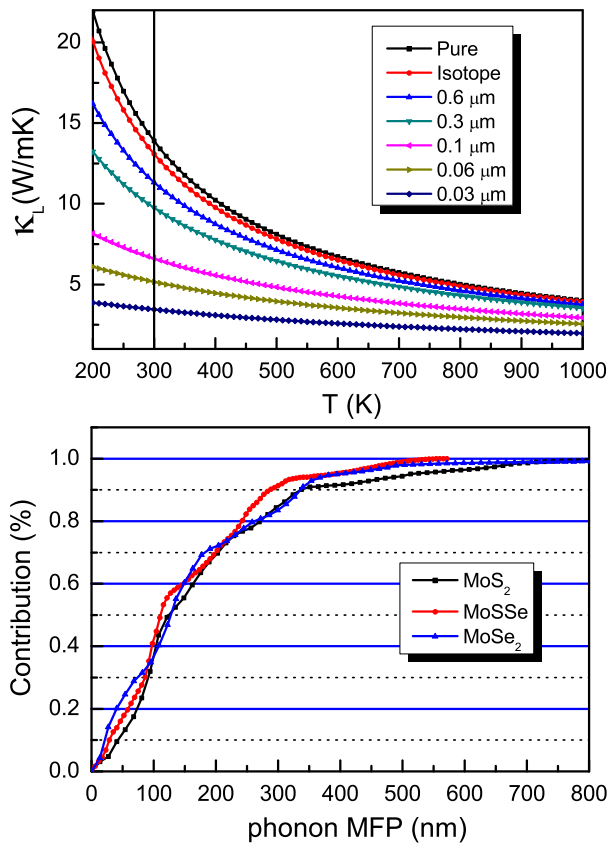}
  \caption{(Color online)Top: the lattice thermal conductivities  of infinite (Pure and Isotope) and finite-size (0.6, 0.3, 0.1, 0.06 and 0.03 $\mathrm{\mu m}$) monolayer MoSSe as a function of temperature; Bottom: the cumulative lattice thermal conductivity of $\mathrm{MoS_2}$, MoSSe and $\mathrm{MoSe_2}$ monolayers  divided by their total lattice thermal conductivity with respect to phonon MFP at room temperature.}\label{mkl}
\end{figure}

To further understand phonon transports of $\mathrm{MoS_2}$, MoSSe and $\mathrm{MoSe_2}$ monolayers,  phonon mode group velocities and lifetimes are calculated.
Due to dominant contribution to total $\kappa_L$ from acoustic phonon branches, we only show  acoustic  phonon mode group velocities and lifetimes in \autoref{v} and \autoref{t}. From $\mathrm{MoS_2}$ to  MoSSe to  $\mathrm{MoSe_2}$ monolayer, most of group velocities  become small  due to softened acoustic phonon modes, which results in the decrease of the lattice thermal conductivity.  The largest phonon group velocity at the
$\Gamma$ point of the LA/TA modes decreases  from 6.60/4.11 km/s to
5.62/3.55 km/s to  5.06/3.22 km/s  from $\mathrm{MoS_2}$ to  MoSSe to  $\mathrm{MoSe_2}$ monolayer.
For ZA branch, the  largest phonon group velocity changes from 4.18 km/s to  3.56 km/s  to  3.29 km/s.
Therefore, the group velocity reduction may be partial reason for the thermal conductivity reduction  from $\mathrm{MoS_2}$ to  MoSSe to  $\mathrm{MoSe_2}$ monolayer. It is straightforward to find that most of phonon  lifetimes  of  MoSSe and  $\mathrm{MoSe_2}$ monolayers are shorter than ones of  $\mathrm{MoS_2}$ monolayer, which may be due to larger  acoustic and optical phonon gap.
However, the phonon  lifetimes between  MoSSe and  $\mathrm{MoSe_2}$ monolayers are comparative.
The lower $\kappa_L$ for MoSSe/$\mathrm{MoSe_2}$ than $\mathrm{MoS_2}$ monolayer is due to lower group velocities and shorter lifetimes.
The  $\kappa_L$ of  $\mathrm{MoSe_2}$ is lower than that of MoSSe, which is mainly due to lower group velocities.
\begin{table*}
\centering \caption{For $\mathrm{MoS_2}$,  MoSSe and $\mathrm{MoSe_2}$ monolayers, the elastic constants $C_{ij}$, shear modulus
$G^{2D}$,  Young's modulus $Y^{2D}$ in $\mathrm{Nm^{-1}}$, and Poisson's ratio $\nu$
dimensionless. }\label{tab3}
  \begin{tabular*}{0.96\textwidth}{@{\extracolsep{\fill}}ccccccc}
  \hline\hline
Name& $C_{11}=C_{22}$ &  $C_{12}$& $C_{66}=G^{2D}$&$Y_{[10]}^{2D}=Y_{[01]}^{2D}$& $\nu_{[10]}=\nu_{[01]}$\\\hline\hline
$\mathrm{MoS_2}$&131.7 (138.5\cite{p2}, 130\cite{p8}, 130.3\cite{p9}) &31.2  (31.7\cite{p2}, 32\cite{p8}, 31.0\cite{p9}) &50.3&124.3& 0.24\\\hline
MoSSe&119.3 (126.8\cite{p2})&27.5 (27.4\cite{p2})&45.9&113.0&0.23\\\hline
$\mathrm{MoSe_2}$&115.6 (115.9\cite{p2}, 108\cite{p8}, 110.1\cite{p9})&26.5 (24.0\cite{p2}, 25\cite{p8}, 26.0\cite{p9})&44.6&109.5&0.23\\\hline\hline
\end{tabular*}
\end{table*}

Based on the formula  proposed by Shin-ichiro Tamura\cite{q24}, phonon-isotope scattering is included, and the  mass variance parameters are read from database of the natural abundance data for elements.  The room temperature
"isotope effect" can be measured by $P=(\kappa_{pure}/\kappa_{iso}-1)$.  The calculated value  is 6.2\%,
which means  that phonon-isotope scattering has little effects on $\kappa_L$. With increasing temperature,  isotopic effect on $\kappa_L$ gradually becomes weak
due to enhancement of phonon-phonon scattering. In reality, finite-size sample is usually used in
the device implementation.  By adopting a most simple boundary scattering model,  the boundary scattering rate can be obtained  by $v_g/L$, in which  $v_g$, $L$  are the group velocity and   boundary mean free path (MFP), respectively. The lattice thermal conductivities  of infinite and  finite-size (0.6, 0.3, 0.1, 0.06 and 0.03 $\mathrm{\mu m}$) MoSSe monolayer as a function of temperature are plotted in \autoref{mkl}. It is apparent
that the thermal conductivity decreases with length decreasing, which is  due to enhanced boundary scattering.
For the 0.6, 0.3, 0.1, 0.06 and 0.03 $\mathrm{\mu m}$ cases,  the room-temperature $\kappa_L$ of MoSSe monolayer  is  about 81.5\%, 70.2\%,  47.6\%, 37.1\% and 24.8\%  of  one  of infinite (Pure) case.

The MFP distributions over a wide range of length scales can be measured by  thermal conductivity spectroscopy technique\cite{pl1}.
At 300 K, the  ratio between cumulative  and   total lattice thermal conductivity  of $\mathrm{MoS_2}$, MoSSe and $\mathrm{MoSe_2}$ monolayers  as a function of phonon MFP are shown in \autoref{mkl}, which  measures  how phonons
with different MFP contribute to the total lattice thermal conductivity.
With  MFP increasing, the ratio  approaches one.  When the ratio reaches 99\%, the corresponding   MFP   is 728 nm,  502 nm  and 748 nm from $\mathrm{MoS_2}$ to MoSSe to $\mathrm{MoSe_2}$ monolayer. The critical MFP of MoSSe  is smaller than that of $\mathrm{MoS_2}$ or $\mathrm{MoSe_2}$ monolayer, which is because MoSSe monolayer  contains more element types. It is noted that critical  MFP   significantly depends on strain, which has be found in antimonene, silicene, germanene, and stanene\cite{l102,l101}. With $\kappa_L$  reducing  to half by nanostructures,  the characteristic length changes from  121 nm to 111 nm to 129 nm  from $\mathrm{MoS_2}$ to MoSSe to $\mathrm{MoSe_2}$ monolayer.

The $\kappa_L$ is connected with Young's modulus by the simple relation $\kappa_L\sim \sqrt{E}$\cite{q16}, and the Young's modulus can be attained
from elastic constants. Due to $D_{3h}$ symmetry,  two independent elastic
constants $C_{11}$=$C_{22}$ and $C_{12}$ can be calculated, and the $C_{66}$=($C_{11}$-$C_{12}$)/2.
\autoref{tab3} lists the  elastic constants $C_{ij}$ of $\mathrm{MoS_2}$, MoSSe and $\mathrm{MoSe_2}$ monolayers, and they all satisfy the  Born  criteria of mechanical stability, namely
\begin{equation}\label{e1}
C_{11}>0,~~ C_{66}>0
\end{equation}
The 2D Young¡¯s moduli $Y^{2D}$ in the
Cartesian [10] and [01] directions  and shear modulus $G^{2D}$ are given\cite{ela}
\begin{equation}\label{e1}
Y^{2D}_{[10]}=\frac{C_{11}C_{22}-C_{12}^2}{C_{22}},~~ Y^{2D}_{[01]}=\frac{C_{11}C_{22}-C_{12}^2}{C_{11}}
\end{equation}
\begin{equation}\label{e1}
G^{2D}=C_{66}
\end{equation}
The corresponding Poisson's ratios can be expressed as:
\begin{equation}\label{e1}
\nu^{2D}_{[10]}=\frac{C_{12}}{C_{22}},~~ \nu^{2D}_{[01]}=\frac{C_{12}}{C_{11}}
\end{equation}
According to \autoref{tab3}, the Young's modulus of MoSSe monolayer  is between ones of $\mathrm{MoS_2}$ and $\mathrm{MoSe_2}$ monolayers, and the order of Young's modulus  is identical with  one of their $\kappa_L$. The calculated $C_{ij}$ agree well with previous theoretical values\cite{p2,p8,p9}, which are also listed in \autoref{tab3}. It is found that the MoSSe monolayer is more flexible  than   $\mathrm{MoS_2}$ monolayer due to  smaller Young's
modulus.

 \begin{figure}
  \includegraphics[width=8cm]{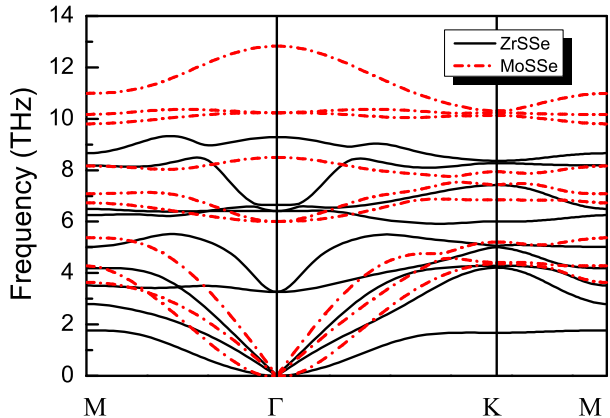}
  \caption{Phonon dispersion curves of ZrSSe and  MoSSe monolayers.}\label{ph-c}
\end{figure}

\section{Discussions and Conclusion}
Recently, the ZrSSe monolayer is predicted by the first principle calculations\cite{p4}, and the calculated room-temperature sheet thermal conductance  is 33.6 $\mathrm{W K^{-1}}$, which is about 9.8\% of one of MoSSe monolayer (342.50  $\mathrm{W K^{-1}}$).  The huge difference on $\kappa_L$ can be understood by their phonon dispersion curves, which are shown in \autoref{ph-c}. It is clearly seen that the dispersion of acoustic branches of ZrSSe monolayer is softened with respect to MoSSe monolayer, indicating the reduction of phonon group velocity, which  leads to lower $\kappa_L$ for ZrSSe than MoSSe monolayer.
The group velocity reduction partially explains the lower $\kappa_L$ for ZrSSe than MoSSe monolayer. A frequency gap
between the optical and acoustic phonon branches in MoSSe monolayer  can be observed, but disappear for ZrSSe monolayer.
The cross between optical and acoustic phonon branches for ZrSSe monolayer leads to  much more frequent
aao scattering, producing very short phonon lifetimes.  The gap for MoSSe monolayer  makes  aao scattering ineffective,  resulting in very long phonon lifetimes. The phonon lifetimes of ZrSSe are almost one order-of-magnitude smaller than that of MoSSe, which can lead to very lower $\kappa_L$ for ZrSSe than MoSSe monolayer.

Strain effects on $\kappa_L$ of  various 2D materials have been investigated\cite{l10,l100,l101,l4-3}. For penta-$\mathrm{SiN_2}$, a  planar structure can be achieved  from a buckled structure by tensile strain,  and the $\kappa_L$ jumps up by 1 order of magnitude\cite{l10}, which is because the reflection symmetry selection rule strongly restricts anharmonic phonon scattering. For penta-$\mathrm{SiC_2}$, the $\kappa_L$ exhibits an unusual nonmonotonic up-and-down behavior\cite{l10}. For $\mathrm{MoTe_2}$, the $\kappa_L$ shows monotonic reduction due to the reduction in phonon
group velocities and phonon lifetime\cite{l4-3}. Therefore, it is very interesting to investigate the strain influence on $\kappa_L$ of MoSSe monolayer.

In summary, based on phonon Boltzmann equation within
the single-mode RTA,   the $\kappa_L$ of MoSSe monolayer is investigated together with  $\mathrm{MoS_2}$ and $\mathrm{MoSe_2}$ monolayers.
Calculated results show that the $\kappa_L$ of MoSSe monolayer is very lower than that  of  $\mathrm{MoS_2}$ monolayer, which is due to the smaller group velocities and  shorter phonon lifetimes for MoSSe  than $\mathrm{MoS_2}$ monolayer. However, the $\kappa_L$ of MoSSe monolayer is higher than that  of  $\mathrm{MoSe_2}$ monolayer, which is mainly  due to larger  group velocities.
It is expected that the order of Young's modulus is  $\mathrm{MoS_2}$ $>$ MoSSe $>$ $\mathrm{MoSe_2}$, which is identical with that of $\kappa_L$.
The isotope effect and  size dependence of $\kappa_L$ of MoSSe monolayer are also  investigated,   which is useful for  designing nanostructures.
This work presents  comprehensive investigations
on the phonon transport of Janus monolayer MoSSe, which is useful  for further study  in TMD Janus monolayers.

\begin{acknowledgments}
This work is supported by the National Natural Science Foundation of China (Grant No.11404391). We are grateful to the Advanced Analysis and Computation Center of CUMT for the award of CPU hours to accomplish this work.
\end{acknowledgments}

\end{document}